# Some comments on the matching of photometric and magnetic properties of structures at the solar surface

*Obridko V.N.[1], Sokoloff D.D.[1,2], Katsova M.M.[3]*

[1] IZMIRAN, 4, Kaluzhskoe Shosse, Troitsk, Moscow 108840, Russia; obridko@mail.ru
[2] Department of Physics, Moscow State University, 119991, Russia; sokoloff.dd@gmail.com
[3] Sternberg Astronomical Institute, Lomonosov Moscow State University, 13, Universitetsky Prospekt, Moscow 119234; mkatsova@mail.ru

**Abstract**

We investigate sharply outlined features recorded in solar magnetic field tracers. It is shown that the magnetic boundaries of a sunspot do not coincide with the photometric ones. Moreover, there is no clear magnetic boundary around sunspots. Thus, the widely accepted concept of a magnetic tube with clearly pronounced borders is not always correct and should be used with caution. It is also shown that even in the periods of complete absence of visible spots on the Sun, there are magnetic fields over 800 Gauss. The nature of these strong magnetic fields remains unclear; they may originate at relatively small depths under the photosphere.

**Introduction**

Since it is known that the photometric and magnetic boundaries of sunspots do not coincide, Obridko et al. (2022) investigated sharply outlined features revealed in solar magnetic field tracers. Taking into account that the magnetic boundary is not determined accurately enough, they showed that photometric and magnetic properties of objects on the solar surface need further investigation. In this paper, we continue discussing the correspondence between the photometric and the magnetic boundaries. It should be noted right away that, generally speaking, the concept of a magnetic boundary is rather conventional. The magnetic field has no walls; so, it is impossible to imagine a physical object on the Sun, inside which the magnetic field exists and outside it is completely absent. This is a fundamental distinction between the magnetic and the photometric boundary, at least at the level of the photosphere, where the horizontal optical thickness is comparable to the resolution limits of up-to-date observational facilities. Therefore, the photometric boundary can be established more reliably than the magnetic one. A different matter is that we can specify the value of the magnetic field, which corresponds to the given photometric boundary.

There is no doubt that the magnetic field is the main factor determining the very existence of most objects in the Sun. However, the definition of their boundaries in terms of the magnetic field is not well developed in scientific literature. The fact is that we can only obtain information about the structure and dynamics of these objects from the analysis of their radiation. The magnetic field does not radiate directly. Radiation comes from plasma, whose characteristics (pressure, density, and temperature) are determined by processes that are affected by the magnetic field.

The magnetic field in the corona is often described as tubes and loops. Indeed, high-quality images of the corona display structures delineated by the field lines. Moreover, extrapolations of the potential field often agree well with the coronal structure (Aschwanden, 2001, 2004; Badalyan, 2013; Mac Cormack, 2022). However, one should remember that the radiation in this case depends on the differential emission measure calculated as a full integral of the squared electron density $n_e^2$ at the corresponding temperature; e.g., see (Priest, 1982; Aschwanden, 2004; Pevtsov, 2003). The density and temperature are determined by heating mechanisms (Parker, 1988; Aschwanden, 2004). The latter can be canalized along the field lines and, in any case, they strongly depend on the magnitude and structure of the magnetic field; see (Badalyan and Obridko, 2006, 2007) and references therein. In this case, weak variations in the magnetic field can significantly affect the radiation of the coronal plasma. The field structural features in the lower corona or in the photosphere can manifest themselves in the particular details of radiation of the coronal plasma. No wonder that in this case, the glow of the solar corona can outline the structure of field lines. The observed photometric feature can be associated with moderate variations in the magnetic field, while the large-scale field as a whole remains quasi-homogeneous. Therefore, identifying a luminous coronal feature with a magnetic tube or loop and assuming a correspondence between their spatial parameters can entail serious errors in modeling and calculation of mechanisms. To date, this question remains open due to the unclear mechanisms of heating and energy transfer in the corona, as well as due to a possible presence of strong currents.

**Magnetic Field at the Photometric Boundary of a Sunspot**

For a long time, the magnetic field outside sunspots was considered negligible. Equations were derived, according to which the magnetic field vanishes at the outer boundary of the penumbra (Broxon, 1942; Mattig, 1953), and the dependence of the field intensity on the distance from the center of a symmetric spot is fully determined by the maximum magnetic field at the center.

The magnetic and photometric boundaries were assumed to coincide. Although this assumption still needs verification, it, nevertheless, resulted in a theoretical concept of magnetic tubes and ropes. Nowadays, the concept of a floating magnetic tube is widely accepted. It is believed, that sunspots arise during the formation of active regions on the solar surface from a strong toroidal field generated by the solar dynamo. In fact, all arguments in favor of this concept are based on theoretical considerations (Caligari et al, 1995, 1998; Fan, 2008; Fan and Fang, 2014; Weber, 2011; Getling, 2016; Getling and Buchnev, 2019). The mechanism of formation of sunspots and (more broadly) bipolar active regions described above has been recently analyzed in detail in (Kosovichev, 2009, 2012; Smirnova et al, 2013; Solov'ev and Kirichek, 2014; Rempel and Cheung, 2014; Zagainova et al., 2022; Getling, 2016; Getling and Buchnev, 2019).

In (Obridko et al., 2022), we proposed a new method for determining the magnetic boundary of visible sunspots. The method was based on long-term data series, including SDO/HMI data on the daily longitudinal magnetic field for 2375 days (from 01.05.2010 to 31.10.2016) and the daily sunspot numbers from WDC–SILSO, Royal Observatory of Belgium, Brussels https://sidc.oma.be/SILSO/datafiles (version 2). The cumulative daily sunspot areas were taken from the NASA Web site https://solarscience.msfc.nasa.gov/greenwch.shtml. At present, two databases of high–resolution observations carried out with single–type instruments are available. These are SOHO/MDI and SDO/HMI. The Michelson Doppler Imager (MDI) onboard the Solar and Heliospheric Observatory (SOHO) was continuously measuring the Doppler velocity, longitudinal magnetic field, and brightness of the Sun for 15 years up to 12 April 2011. The enhanced Helioseismic and Magnetic Imager HMI onboard the Solar Dynamics Observatory (SDO) started its routine observations on 30 April 2010. HMI observations provide the same data as MDI, but with much higher spatial and temporal resolution and of better quality.

We estimate the relative area of the solar surface occupied by the magnetic field exceeding a certain threshold value. This area is expressed in millionths of the visible hemisphere (m.v.h.) as is customary in studying the total sunspot areas. Then, we compare these calculations with the daily sunspot data to find the field value, at which the area of the closed region bounded by this field according to magnetic measurements corresponds exactly to the sunspot photometric area on that day. As a result, we arrive at the conclusion that the magnetic field of a sunspot as defined by the field normal component is about 550 G.

However, the magnetic field at the visible boundary of a spot does not vanish. The field region does not have a sharp boundary and extends far beyond the photometric boundary. The relationship between the field intensity in a circular sunspot and its relative photometric radius is illustrated in Figure 1.

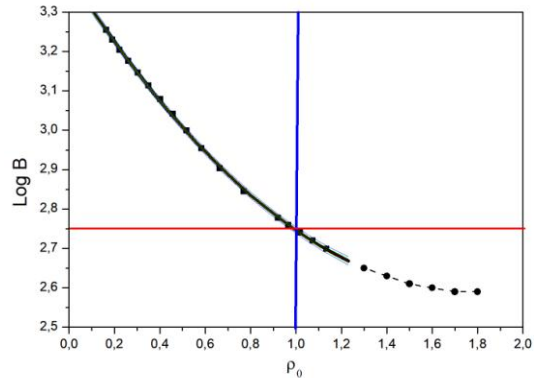

**Figure 1.** Empirical dependence of the magnetic field (dots, measured in G) in a symmetric sunspot on the relative sunspot radius ($\rho_0 = 1$ at the photometric boundary). The solid line is a polynomial approximation, the dashes extrapolate the approximation for weak magnetic fields. The thin blue lines near the approximation mark the 95% confidence interval. The thick blue line shows the photometric sunspot boundary, and the red line shows the magnetic sunspot radius, calculated presuming that the sunspot magnetic radius corresponds to the magnetic field strength of 550 G.

Thus, the sharpness of the boundary is not the result of the absence of magnetic field outside the "tube", but is rather due to a combined action of two interrelated phenomena. At a certain field value (i.e., at 500-550 G), the convective transfer from depth to the photosphere is suppressed, and the temperature (and, hence, the brightness of the region) drops. The energy of the 500 G magnetic field is comparable to that of the convective flow in the lower part of the photosphere at the optical depth $\tau = 1$ at a density of about $3 \cdot 10^{-7}$ g/cm$^{-3}$ and a speed of about 2 km/s. In this case, the sharpness of the boundary is determined by the fact that the horizontal optical thickness in the photosphere does not exceed 100 km and, therefore, the horizontal heating for the sunspot as a whole is ineffective.

**Small Elements with a Strong Magnetic Field**

The discussion above is aimed at determining the boundary of large spots. However, there is a local aspect of the problem. The fact is that the interaction between the magnetic field and the convective transfer depends on the size of the magnetic element. If the element is small enough (about 100 km in size), the horizontal optical thickness for the radiative transfer becomes comparable to the geometric size of the element. Then, the horizontal transfer can smooth out the temperature profile. This is why some elements with a strong magnetic field are present on the solar surface even if standard observations do not reveal any sunspots.

Stenflo (1973, 1982) claimed the existence of optically unobservable small elements of the magnetic field. Shiota et al. (2012) directly observed such elements in the solar polar region with extra-atmospheric high-resolution instruments.

In order to estimate the role of locally strong magnetic fields in the context of sunspot studies, we illustrate observations of elements with locally strong magnetic field obtained on spotless days (Figure 2). We have calculated the area covered with the field above the given threshold value.

The black dots stand for the magnetic field B>100 125 G. The area of these regions is of the order of several thousand m.v.h. The areas of the regions covered with magnetic fields B>400 G (squares) and B>600 G (circles) are slightly smaller.

However, it is most impressive that, even on spotless days, there are dozens of objects with B>800 G (crosses). Such objects should be considered as sunspots, although they are not observed by standard methods. The area of such regions is very small (several dozen m.v.h.), as well as their contribution to the total magnetic flux. Still the total magnetic field energy in these regions may reach $10^{30}$ erg and they may be responsible for moderate solar flares. We want to emphasize that dozens of such objects exist even in the epochs of very deep solar minimum. As they are not recorded by the sunspot patrol service, their size is apparently much smaller than 2-3 arcsec in cross section.

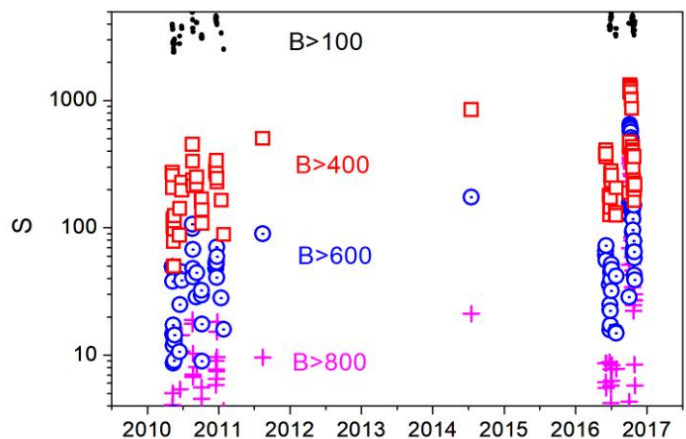

**Figure 2.** Statistics of the magnetic-field intensity on spotless days. The black dots stand for B>100 G, the red squares stand for B>400 G, the blue dotted circles correspond to B>600 G, and the purple crosses indicate the area covered by strong magnetic field B>800 G.

The role of small magnetic elements has to be somehow included in the scenario of the solar cycle. According to the conventional scenario, the large-scale toroidal magnetic field is produced from the poloidal field by differential rotation. The small magnetic elements are not inevitably formed as part of this scheme, but can be driven by turbulent processes. During the solar minima, i.e., in the periods, when the action of the large-scale dynamo is weak, turbulent mechanisms may produce small magnetic elements all over the solar surface and not just near the solar equator (see Sherrer et al,. 2012). Note that at the reversal of the large-scale polar magnetic field, the number of small magnetic elements with strong magnetic fields of both polarities is more or less equal.

**Conclusion**

Thus, we arrive at the conclusion that the sunspot is not an isolated magnetic tube. The photometric boundary of a sunspot is not its physical boundary, but is the region of a sharp change in the spot brightness as the magnetic field reaches the threshold value of 550 G. On the other hand, small elements with high magnetic fields permanently exist in the Sun, but their contrast is too low due to the horizontal heat transfer, so that standard observations cannot identify them as spots. The brightness drops significantly only in large features that significantly exceed the horizontal optical thickness (≥100 km.)

The results obtained are of great importance for understanding the generation of magnetic field on the Sun and the formation of active regions. The generally accepted idea of the field tubes is not entirely correct. It is believed that sunspots appear as individual features on the solar surface in the course of the formation of AR from a strong toroidal field, which is generated by the solar dynamo mechanism at the base of the convection zone and is carried out into the photosphere. In fact, the arguments in favor of this concept are based rather on theoretical considerations than on reliable observational evidence. The emergence of a single magnetic tube as a source of sunspots contradicts the observed field structure of a single sunspot. During the generation process, the turbulent dynamo creates many elements with different field strengths. Their energy distribution changes with the phase of the cycle. However, these elements are not tubes with isolated boundaries. The field in them gradually decreases with distance from the center of the element to its periphery. The photometrically sharp boundaries of the spots are the result of influence of the magnetic field on the energy transfer process. Fields above 550 G greatly reduce the energy flux from below and a sharp boundary appears (Pikelner, 1960; Kaplan and Pikelner, 1974).

Moreover, sunspots emerge in the pre-existing magnetic environment and are included in active regions. The formation of sunspots is not at all a surface phenomenon. It rather develops in the leptocline and obviously requires further investigation and modeling (see, e.g., Kitiashvili, 2023).

Note that our analysis of objects on the solar surface, probably, entails a more general conclusion: the existence of objects with sharp photometric boundaries both on the Sun and in space by no means implies the existence of as sharply bounded magnetic structures.

**Acknowledgements**

VNO, MMK and DDS acknowledge the support of the Ministry of Science and Higher Education of the Russian Federation under the grant 075-15-2020- 780 (VNO and MMK) and 075-15-2022-284 (DDS). DDS thanks support by BASIS fund number 21-1-1-4-1.